\documentstyle[aps,prl,multicol,fancyheadings,epsfig,amsbsy,amssymb,amstex]{revtex}  

\def\bbbc{{\mathchoice {\setbox0=\hbox{$\displaystyle\rm C$}\hbox{\hbox 
to0pt{\kern0.4\wd0\vrule height0.9\ht0\hss}\box0}} 
{\setbox0=\hbox{$\textstyle\rm C$}\hbox{\hbox 
to0pt{\kern0.4\wd0\vrule height0.9\ht0\hss}\box0}} 
{\setbox0=\hbox{$\scriptstyle\rm C$}\hbox{\hbox 
to0pt{\kern0.4\wd0\vrule height0.9\ht0\hss}\box0}} 
{\setbox0=\hbox{$\scriptscriptstyle\rm C$}\hbox{\hbox 
to0pt{\kern0.4\wd0\vrule height0.9\ht0\hss}\box0}}}}

\begin{document} 

\draft 
\title{Charge and spin inhomogeneous phases in the Ferromagnetic Kondo Lattice Model} 
\author{D. J. Garcia,$^1$ K. Hallberg,$^1$, C. D. Batista,$^2$
S. Capponi,$^3$ D. Poilblanc $^3$, M. Avignon $^4$ and B. Alascio $^1$} 
\address{$^1$Centro At\'omico Bariloche and Instituto Balseiro\\ 
Comisi\'on Nacional de Energ\'{\i}a Atomica, 8400 S. C. de Bariloche, Argentina,\\$^2$Center for Nonlinear Studies and Theoretical Division\\ 
Los Alamos National Laboratory, Los Alamos, NM 87545\\
$^3$ Laboratoire de Physique Quantique IRSAMC\\ Universite Paul Sabatier, 118 Route de Narbonne, 31 062, Toulouse, France\\
$^4$ Laboratoire d'Etudes des Propriétés Electroniques des Solides-CNRS,
Associated with Université Joseph Fourier, BP 166, 38042 Grenoble, Cedex 
9, France}
\date{\today} 
\maketitle 
\begin{abstract} 
We study numerically the one-dimensional ferromagnetic Kondo lattice. This model is widely used 
to describe nickel and manganese perovskites. Due to the competition between double and 
super-exchange, we find a region where the formation of magnetic polarons induces a 
charge-ordered state. This ordering is present even in the absence of any inter-site 
Coulomb repulsion. There is an insulating gap associated to the charge structure formation. 
We also study the insulator-metal transition induced by a magnetic field which removes 
simultaneously both charge and spin ordering. 
\end{abstract} 
 
\begin{multicols}{2} 
 
\columnseprule 0pt 
 
\narrowtext

\section{Introduction}

A new paradigm in condensed matter springs up as a consequence of 
the growing body of experimental evidence suggesting different types 
of inhomogenous states in strongly correlated systems with competitive 
interactions. The simultaneous appearence of charge and spin 
superstructures seems to be present in completely different systems. 
For example, charge, lattice and spin modulations with a doping-dependent
wave vector have been observed in La\( _{2-x} \)Sr\( _{x} \)NiO\( _{4} \) 
by means of neutron scattering\cite{hayden} and electron diffraction\cite{chen}.
Stripe formation, together with incommensurate spin fluctuations in high Tc 
superconductors can also be regarded as a manifestation of similar phenomena\cite{tranq}. 
The charge and spin ordering found in many of the doped manganese perovskites fall 
in the same category. Experiments have revealed charge ordering (CO) at half filling 
in Nd\(_{0.5}\)Sr\(_{0.5} \)MnO\(_{3} \)\cite{kawano} and similar compounds, 
such as Pr\( _{0.5} \)Ca\( _{0.5} \)MnO\( _{3}\),\cite{tomioka,li,moritomo,chen2}.
More recent interest has focused on electron doped charge ordered manganites.
The first observation of charge, orbital, and magnetic ordering in electron 
doped manganites dates back to the work of Wollan and Koehler \cite{wollan} 
on La\( _{1-x} \)Ca\( _{x} \)Mn O\( _{3} \). Several 
works confirmed the observation of this behavior in other 
compounds \cite{liubao,mori,ueharajirac}.
CO has also been found for other dopings as 
in Bi\( _{1-x} \)Ca\( _{x} \)MnO\( _{3}\)\cite{liubao} 
and in La\( _{1-x} \)Ca\( _{x} \)MnO\( _{3} \) 
(doped with \( P_{r} \)) \cite{ueharajirac} for \( x\geq \frac{1}{2} \) (few electron region). 

Manganese oxides have been intensively studied during the last years due to their collosal 
magnetoresistance (CMR) effect, i.e. a large change in the resistivity accompained by a 
transition from an insulating antiferromagnetic (AF) or paramagnetic state to a ferromagnetic 
(FM) metal. In the above mentioned cases, the AF state also exhibits charge ordering, and
the insulating gap is associated to the formation of the charge structure.

Numerical studies for several dopings\cite{malvezzi} have included the 
effect of nearest neighbor density-density repulsion \( V \) and obtained a very rich phase diagram. 
The diagram includes phase separation
for either extreme dopings ($x\simeq 0$ or $x\simeq 1$), and CO for concentration $x\simeq 0.5$. 
Using Monte Carlo techniques, Yunoki {\it et. al.} \cite{yunoki} studied the half filled case (\( x=0.5 \))
including both $e_g$ orbitals.
They found that the CO phase is stabilized by Jahn-Teller phonons. However, as it is mentioned in their work \cite{yunoki}, 
the \( z \)-axis stacking of charge and the existence of bistripes at \( x>0.5 \)\cite{mori}
are both penalized by a large Coulomb interaction. This observation indicates that the Coulomb repulsion \( V \) 
is smaller than the value which is required to drive a charge ordered state. 

Unlike non-magnetic materials, where the Peierls instability or large inter-site Coulomb interactions are required
to get CO, we show here that charge density waves can result in manganites from the formation of 
magnetic superstructures arising from the presence of competing interactions. A very rich family of inhomogeneous spin and charge
structures \cite{daniel} can be obtained with this mechanism, by changing the carrier concentration 
or the ratio between the competing interactions.

In order to illustrate our new mechanism, we will study here a simplified model where the 
competing forces are represented by the double and super-exchange interactions.
The Ferromagnetic Kondo Lattice Model (FKLM), proposed to describe the manganites, 
was studied originally by de Gennes\cite{de gennes}. One of the most intriguing questions related to
this model is: what are the emerging phases when the competing 
interactions are energetically similar? De Gennes proposed canting of two interpenetrating lattices as the compromise 
solution for this competition. This concept was used subsequently in recent analytical approaches \cite{Aliaga,Arovas}. 
In addition,  
phase separation has also been considered as a possible solution to this competition \cite{Kagan,Moreo}. 
In contrast, in the present paper
we find spin phases which cannot be described in terms of two interpenetrating sublattices nor 
do they correspond to phase separation. 

As a result of the competition between the double exchange (DE) mechanism which delocalizes the hole,
and the super-exchange (SE) between local spins, different phases may appear such as ferromagnetism in one extreme, 
antiferromagnetism in the other, and doping-dependent modulated charge and spin order in between.
For \( x=1/2, \) we do not find charge ordering for \( V=0 \). However for concentrations which are away from half filling,
we obtain charge modulation even in the absence of Coulomb interactions. This new state can be described 
as an ordered phase of FM islands where carriers are localized.

Since the charge modulation is driven by the AF ordering between the islands, 
the transition to a FM state destroys the charge ordering, closing the insulating gap.
This transition can be induced by applying an external magnetic field. If the energy difference 
between the AF-charge ordered  phase and the FM state is small, the transition is induced by a magnetic field
which can be generated in the laboratory. In this respect, this new mechanism gives a natural 
way to understand how the interplay between charge and 
spin degrees of freedom can induce a collosal magnetoresistive effect. 
The complete understanding of this phenomena in 
manganites requires a more detailed study, exploring regions of concentration where there is no charge 
ordering since the concentration of particles is not commensurated with the lattice. 

To study the insulator-metal transition 
we have computed the Drude weight by exact diagonalization of 
finite size systems.
The phase diagram has been computed using the Density Matrix Renormalization Group (DMRG) \cite{white,Karen}. The energy of the different phases has been computed exactly for the case where the localized spins are classical. In this way we compare the phase diagrams correponding to the classical and quantum limits of the spin variables.

In the next section we introduce the model. In Section III, we study the phase diagram making a detailed description of its different  phases and the way they have been obtained. In Section IV we study the charge excitations by calculating the Drude weight, the charge gap, and the
effect of a magnetic field on the metal-insulator transition. Section V is devoted to the conclusions.

\section{The model}

In the manganese oxide compounds, the Mn$^{4+}$ ions consist of three localized $t^3_{2g}$ 
orbitals that can be represented by an $S=3/2$ spin due to the large Hund interaction $J_H$. 
The Mn$^{3+}$ ions have an additional electron in the doubly degenerated $e_g$ orbital 
(where the degeneracy is lifted due to Jahn-Teller interactions).  
To describe this system we use the so called  Ferromagnetic Kondo Lattice Model (FKLM) 
\cite{riera,yunok,yuno1}. This 
Hamiltonian includes a lattice of localized spins which are coupled via $J_H$ to 
itinerant $e_g$ electrons:

\begin{eqnarray}
\label{Hamiltoniano}
H&=&-t\! \sum_{<ij>\sigma} (c_{i\sigma}^{\dagger }c_{j\sigma}+c_{j\sigma}^{\dagger }c_{i\sigma})
+U\! \sum_{i}
n_{i\uparrow
}n_{i\downarrow
}+V\! \sum_{<ij>} n_{i}n_{j}\nonumber \\ 
& & -J_{H}\! \sum_{i} {\bf S}_{i} {\bf \sigma}_{i}+K\! \sum_{<ij>} {\bf S}_{i} {\bf S}_{j}
\end{eqnarray}
where $n_{i\sigma}=c_{i\sigma}^{\dagger }c_{i\sigma}$ and $n_i=n_{i\uparrow}+n_{i\downarrow}$.
The first term represents the $e_{g}$-electron hopping between nearest-neighbor Mn 
ions at sites $i$ and $j$. The second and third terms are the on-site and nearest-neighbor Coulomb repulsions 
for the same orbitals. \( J_{H} \) is the Hund's coupling between localized \( S_{i} \), 
and itinerant \( \sigma _{i} \) spins. We will denote the concentration of holes by $x$, i.e.
$(1-x)$ is the concentration of itinerant electrons.
Hund's interaction together with the hopping term give rise to the DE interaction which favors FM 
ordering of the local spins. 
$K$ is the antiferromagnetic  SE interaction between local spins ($t^3_{2g}$ electrons). This term stabilizes 
the AF phase for $x=1$, and competes with the DE term for intermediate fillings.
For completeness we add the effect of a nearest-neighbor repulsion $V$ when considering \( x=0.5 \). 
The localized spins are taken as $S=1/2$, instead of $S=3/2$, to simplify the calculations. The model is studied
in one dimension (1D), since the basic features of the phase diagram for the 3D case
appear in one and two dimensions as well \cite{yunok,Martin-Mayor}.

Experiments indicate that \( J_{H}\gg t \). For this reason  we fix the values of  \( J_{H}=20t \) and $U=10t$.
Then, the only free parameters are $V$ and $K$. For the large value of  \( J_{H} \) used in this paper, the results 
are not sensitive to \( U \) since double occupancy is suppressed by
\(J_{H}\).
For this reason, we propose to use a t-J model for the conduction electrons hence reducing the Hilbert space by eliminating the doubly occupied states:

\begin{eqnarray}
{\tilde H} & = & -t\! \sum_{<ij>\sigma} ({\bar c}_{i\sigma}^{\dagger
} {\bar c}_{j\sigma}+{\bar c}_{j\sigma}^{\dagger
} {\bar c}_{i\sigma})+J\! \sum_{<ij>} {\bf \sigma}
_{i}{\bf \sigma}
_{j}\label{hamReducido} \\
 &  & +V\! \sum_{<ij>} n_{i}n_{j}-J_{H}\! \sum_{i} {\bf S}_{i} {\bf \sigma}_{i}+K\! \sum_{<ij>}
{\bf S}_{i} {\bf S}_{j} \nonumber
\end{eqnarray}
where 
${\bar c}^{\dagger}_{i\sigma}=c^{\dagger}_{i\sigma}(1-n_{i {\bar \sigma}})$ and
${\bar c}^{\;}_{i\sigma}=(1-n_{i {\bar \sigma}})c^{\dagger}_{i\sigma}$, are the creation and anihilation operators of constrained fermions (no more than one electron per site). 
Strictly speaking, the exchange between two $e_g$ electrons depends on the configuration of the localized spins, involving a more complicated form of the reduced Hamiltonian. However, we are concerned with the low-energy properties so that simply assuming the above form~(\ref{hamReducido}) with $J=\frac{4t^{2}}{U+J_{H}}$ is sufficient.
From the expression for $J$, we see that the condition for the validity of ${\tilde H}$ as a low energy effective theory for $H$ is $t \ll U+J_H$.
We have checked on small systems that ${\tilde H}$ gives similar results as the full Hamiltonian $H$.   
We will use ${\tilde H}$ to calculate the Drude peak and the charge gap (see Sec. IV).

These models are solved using exact diagonalization (Lanczos) and DMRG techniques in open systems. 
When we used DMRG, we took special care in the growing procedure to avoid lattice frustration. 
For instance, we considered different right and left blocks.
We obtained maximum discarded weights of the order of $10^{-5}$ using the finite-size 
algorithm and keeping 150 states. To characterize the different phases we 
calculated static correlation functions for system sizes up to $L=40$ sites. 
This is described in the following section.

\section{Phase diagram}

The charge and spin correlation functions are computed for the complete Hamiltonian $H$ (Eq. (1)),
using DMRG and open boundary conditions. These correlation functions are the spin and charge 
structure factors:

\begin{equation}
\label{eqSq}
S(q)=\frac{1}{L}\sum _{i,j}e^{i(R_{j}-R_{i})q}\left\langle {\bf
S}_{i}{\bf S}_{j}\right\rangle 
\end{equation}
 
\begin{equation}
\label{eqNq}
N(q)=\frac{1}{L}\sum _{i,j}e^{i(R_{j}-R_{i})q}\left\langle (n_{i}-n)(n_{j}-n)\right\rangle,
\end{equation}
where $n=1-x$ is the average number of electrons per site,  $L$ is the chain length, and $S_{i}$ is the localized 
spin operator at site $i$. Both correlation functions are used to build the phase diagram of $H$. 
We describe below the different phases obtained in this way. 

\subsection{\textit{Ferromagnetic phase (FM)}}

\begin{figure}
\begin{center}
\includegraphics[height=6cm]{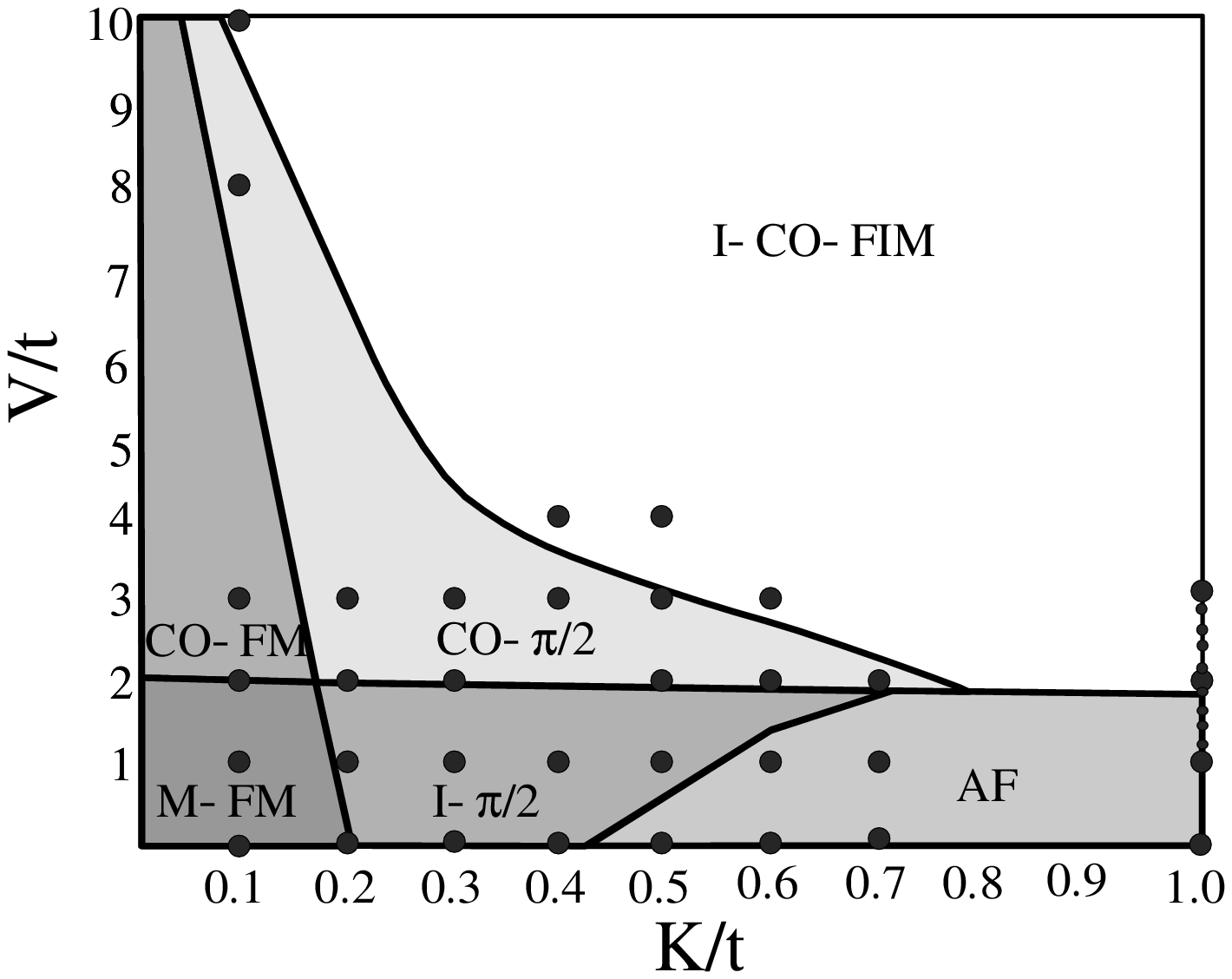}
\end{center}
\caption{Phase diagram for $x=1/2$. The circles indicate the
points where calculations have been performed. CO, I and M stand for
charge order, insulating and metallic phases respectively. FM, FIM, AF
and $\pi /2$ stand for ferro-, ferri-, antiferromagnetic and $q=\pi /2$
spin phases respectively.\label{phasediagram}}
\end{figure}

When \( K\ll t \) the local spins are aligned in a FM 
state (Figs. \ref{phasediagram}, \ref{corrspin}a). At the cost of the AF interaction, the system gains kinetic energy when neighboring spins are parallel. In this situation, the itinerant 
electrons can be mapped into spinless fermions because the large Hund's interaction 
and the FM ordering of the localized spins force the  band electrons to be fully polarized.

%

The charge correlation function for the spinless model can be calculated analytically. It has the shape 
of a truncated pyramid. The truncation begins at \( q=2k_{F} \), where for the spinless case
\(k_{F}=\pi (1-x)\). 
This is shown in Fig. \ref{corrcarga} for $x=1/2$. The oscillations are due to the open boundaries conditions. 
For this case, a charge ordered phase sets in when $V$ reaches the critical value $V_{c}=2t$. There
is a region above this critical value where the spins remain ferromagnetically ordered. However, larger values 
of $V$ inhibit the DE mechanism by localizing the charges, and other spin orderings arise (Fig. \ref{phasediagram}).
For $V=0$, our numerical calculations for the fully quantum case (S=1/2) lead to an upper boundary 
of the FM region at K=0.2t.

\subsection{\textit{Island phase (ILP)}}

When the localized spins are considered as classical degrees of freedom, canted or spiral 
states have been proposed \cite{de gennes} for
the transition between the FM ($K/t=0$) and the AF states ($K\gg t$). 
These phases are obtained by variational 
methods which exclude inhomogeneous states as candidates to be the ground state. 
A more complete calculation with classical spins \cite{yamanaka} shows that the $2k_F$ instability 
of itinerant electrons in one dimension 
leads to the type of spin and charge ordering that we find with our numerical calculations with quantum spins. 
Indeed our DMRG results confirm and extend this picture. For the intermediate $K/t$ regime, the spin and 
charge structure factors peak at \( q=k_{F} \) and \( q=2 k_{F} \) respectively ($k_F$ 
corresponds to the spinless case). 

\begin{figure}
\begin{center}
\includegraphics[height=5cm]{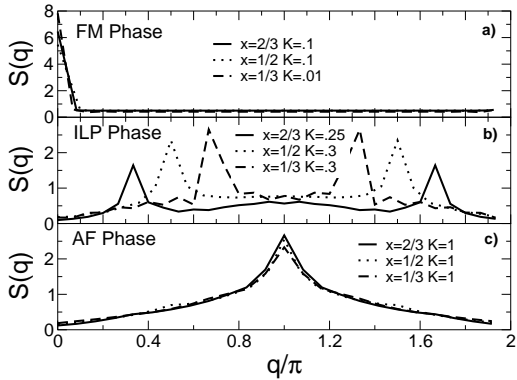}
\end{center}
\caption{Spin correlations for the different phases found ($V=0$): a) Ferromagnetic phase (FM); b) Island-like Phase (ILP); c) Antiferromagnetic Phase (AF) \label{corrspin}}
\end{figure}

\begin{figure}
\begin{center}
\includegraphics[height=5cm]{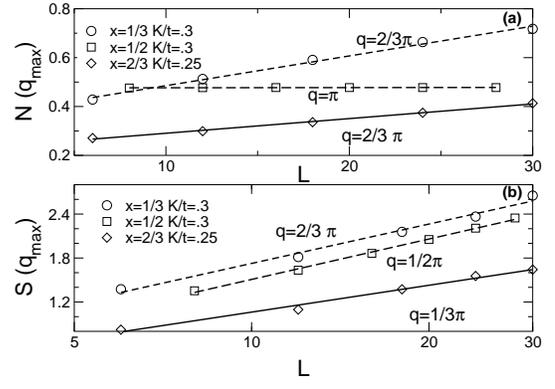}
\end{center}
\caption{a) Charge and b) spin peak-height scalings in the island-type phases. The spin correlation is plotted on a semi-logarithmic scale.\label{peakscaling}}
\end{figure}

The following are the results for some particular fillings:

a) $x=2/3$. In this case the phase roughly consists of FM islands of 3 spins which are aligned antiferromagnetically 
($\uparrow \uparrow \uparrow \downarrow \downarrow \downarrow \uparrow \uparrow \uparrow
\downarrow \downarrow \downarrow$). 
The spin structure factor for the fully quantum case is shown in Fig. \ref{corrspin}b. 
Two well defined peaks are observed at $q=\pm\pi/3$. 
The nearly logarithmic increase of these peaks with the system size $L$ (see Fig. \ref{peakscaling}b) 
indicates a power-law decay with exponent one for the spin-spin correlation as a function of the relative distance. 
However the system sizes considered here do not exclude other possible values for this exponent. 
In any case, the linear increase of the peak in a logarithmic scale indicates that the correlations are quasi long-ranged.
The absence of real long-range order is a consequence of the one-dimensional character of the  
model and the $SU(2)$ continuous symmetry.
The same behavior is observed for the other spin structures described below.

In the classical limit there is no effective hopping between spins which are antiferromagnetically aligned
since $t_{class}=t\cos(\theta/2)$ (where $\theta$ is the relative angle between the spins). Therefore it is clear
that the  charge is then localized 
in each island when the spins are classical. For the quantum case the charges are still localized because the effective hopping between
spins in the same island is still larger than the hopping between spins in different islands. The last
observation plus the fact that there is one electron per island allows to understand the opening of a gap at $q=k_{F}$.

\begin{figure}
\begin{center}
\includegraphics[height=5cm]{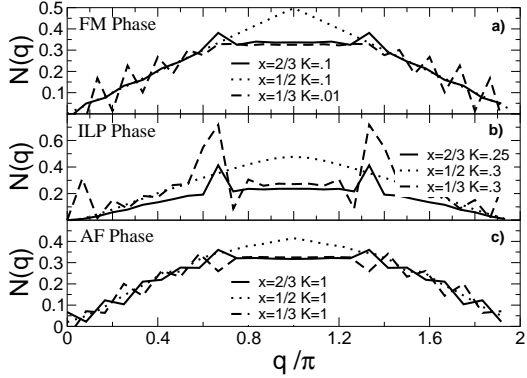}
\end{center}
\caption{Charge correlations for the different phases mentioned in Fig.  1 ($V=0$).\label{corrcarga}}
\end{figure}

\begin{figure}
\begin{center}
\includegraphics[height=6cm]{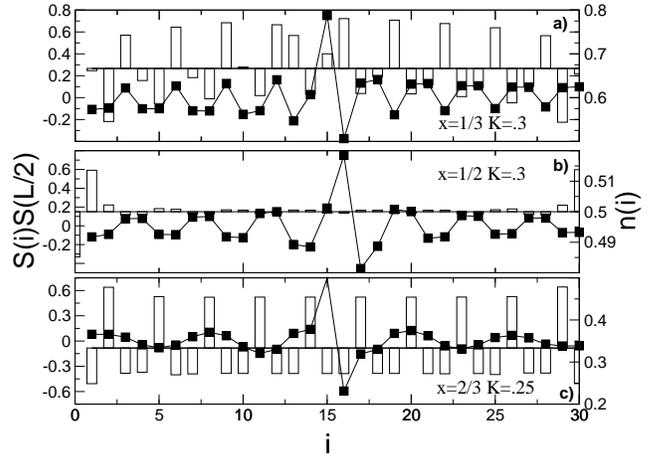}
\end{center}
\caption{Charge mean values (bars, right axis) and spin real space correlations (squares, left axis) for the island type phases. The horizontal line represents the average charge.\label{realspacecorr}}
\end{figure}

The localization of the charge is confirmed by the non-uniform distribution shown in Fig. \ref{realspacecorr}c.
The translation symmetry of $H$ is spontaneously broken in the thermodynamic limit for the charge
ordered state. Even though the symmetry is only broken for the infinite-size chain, the DMRG process selects one of the broken symmetry states also for finite chains.
The charge occupations per site (0.45 in the borders and 0.27 in the center) are very close to the values expected for the classical limit (0.25 in the borders and 0.5 in the center for a three sites open chain). 
Consistently with this picture, the charge-charge structure factor (Fig. \ref{corrcarga}b) shows remarkable peaks for $q=\pm2\pi/3$. Fig. \ref{peakscaling}a shows that the intensity of these peaks scales linearly with $L$.
This is evidence of long-range charge ordering. This behavior is also found for other commensurated fillings like $x=1/3$.

b) $x=1/3$. The classical image of the intermediate $K/t$ phase for this filling is: 
$\uparrow \uparrow \downarrow \uparrow \uparrow \downarrow \uparrow \uparrow \downarrow
\uparrow \uparrow \downarrow$. 
Here we find two-site FM islands separated by an antialigned spin (see Fig. \ref{realspacecorr}a).  
$S(q)$ shows peaks at $q=\pm 2\pi/3$ (see Fig. \ref{corrspin}b). From the 
real space spin-spin correlation function we can rule out the uniform spiral phase since the `islands' are clearly 
distinguishable (this is also valid for $x=2/3$). In a classical image, the charge distribution 
should present one electron in the down spin and another distributed between the two parallel spins.
Our numerical results show a density of approximately 0.6 electrons on the parallel up spins (FM spins) and 0.8 on the anti-aligned spins.
In this case, the inclusion of a hopping between the up and down spins is essential to describe the numerical results. 
$N(q)$ shows the expected peaks at $q=\pm 2\pi/3$ (Fig. \ref{corrcarga}b). 
The peak near zero is due to finite-size effects.

\begin{figure}
\begin{center}
\includegraphics[height=5cm]{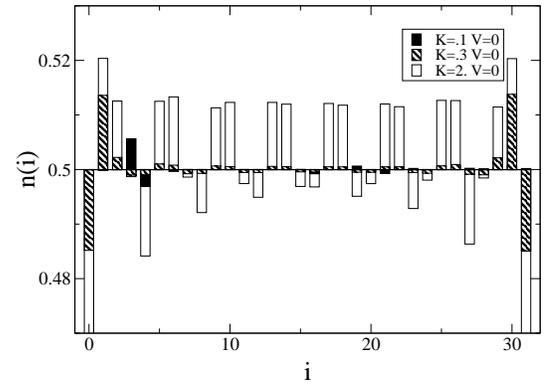}
\end{center}
\caption{On-site charge mean values for x=1/2 in the FM (black), ILP (striped) and AF
(empty) phases.\label{distribcarga1s2}}
\end{figure}

One interesting point is that for classical spins there is electron-hole symmetry; i.e. $x$ and $(1-x)$ 
should exhibit the same phase for a given set of parameters. 
The absence of that symmetry for quantum spins 
is reflected in our results (the ILP for $x=1/3$ is different from the one corresponding to $x=2/3$).

c) $x=1/2$. In this case the FM islands contain two spins and, like in the previous cases, the islands are antiferromagnetically ordered:
($\uparrow \uparrow \downarrow \downarrow \uparrow \uparrow \downarrow \downarrow$).
$S(q)$ shows a clear peak at $q=\pm \pi/2$ 
as it is expected for a unit magnetic cell containing four spins (see Fig. \ref{corrspin}). 
The same structure has been obtained with the Monte Carlo method  for classical
spins\cite{yuno1}. However, this state has been interpreted as a spiral phase.
Other classical Monte Carlo calculations \cite{Horacio} confirm the peak at $q=\pi/2$ but
instead of a spiral phase, they show the island state that we obtain with quantum spins.
 The effective hopping alternates between two values from one bond to the next one in this particular 
 island phase  being larger between the parallel spins. 
 This alternation of the hopping does not induce any charge inhomogeneity, i.e. if $V=0$ there 
 is no charge density wave for this concentration (see Figs. \ref{peakscaling}, \ref{realspacecorr} and 
\ref{distribcarga1s2}). 
However $N(q)$ differs from the FM case. This change is related to 
the presence of bond ordering (dimerization). 
The charges are `localized' in bonds due to the particular spin structure generated by the two competitive interactions. 
It is well known that this type of ordering also opens a charge gap which gives rise to an insulating behavior. We will come back to this in Sec. IV. 
The shape of $N(q)$ is closer to $(1-\cos q)/4$ (charge structure factor of a completely dimerized state) as a consequence of the enhancement of the charge correlations within the islands (bonds).

\begin{figure}
\begin{center}
\includegraphics[height=5cm]{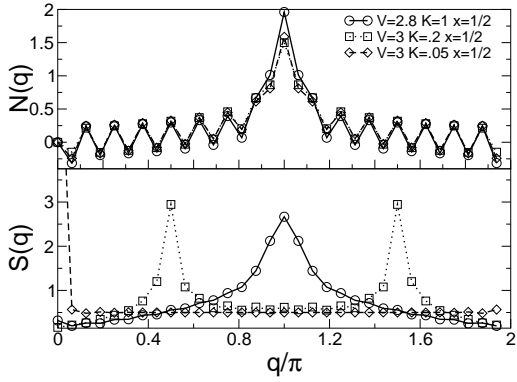}
\end{center}
\caption{Spin and charge correlations for half filling in the presence of nearest neighbor Coulomb interaction.\label{correlationswithV}}
\end{figure}

The above results are for $V=0$. When increasing the Coulomb interaction $V$, a CO phase appears for $V\simeq 2t$ like in the spinless model. 
For $V\gtrsim 2t$, the usual CO phase with a charge structure factor peaked at $q=\pi$ is obtained. $S(q)$ still has a peak at $q=\pi/2$ (Figs. \ref{phasediagram}, \ref{correlationswithV}). Larger values of 
$V$ inhibit the double exchange mechanism since the charge fluctuation are strongly supressed. The magnetic correlations change then from 
$q=\pi/2$ to a ferrimagnetic (FIM) CO phase. This phase can be schematically represented by $\Uparrow \downarrow \Uparrow \downarrow \Uparrow \downarrow $, where the total spin per site is alternating between $S=1$($\Uparrow$) and $S=1/2$ ($\downarrow$). The $S=1$ 
spin at site $i$ corresponds to the local triplet state forced by $J_h$ 
between the itinerant electron and the localized spin at the same site. \\

The same calculations for $x=4/5$ and $3/4$ show spin and charge correlation functions consistent with the formation of five and four FM-spin islands respectively. 
These structures could be regarded as a crystallization of the magnetic polarons described in references \cite{Nos1} for the dilute limit. It is interesting to note that charge ordering is induced by spin ordering and vice versa.
The formation of these superstructures is then a consequence of the interplay between charge and spin degrees of freedom.

\subsection{\textit{Antiferromagnetic phase (AF)}}

As $K$ is increased, the AF ordering prevails over the DE mechanism for any concentration. 
For instance, at half-filling the transition to the AF phase occurs for $K/t \simeq 0.7$ (for $x\neq 1/2$ the transition occurs for smaller values of $K/t$).
$S(q)$ shows a strong peak for $q=\pi$ and the charge is  
uniformly distributed along the sample, i.e. $N(q)$ is broader.
As mentioned above, a FIM charge ordered state is obtained for half filling by increasing $V$. 
(Figs. \ref{phasediagram},\ref{correlationswithV}). 
The charges remain in one sub-lattice forming a chain with alternating S=1 and S=1/2 spins which are antiferromagnetically aligned. \\

\begin{figure}
\begin{center}
\includegraphics[height=5.5cm]{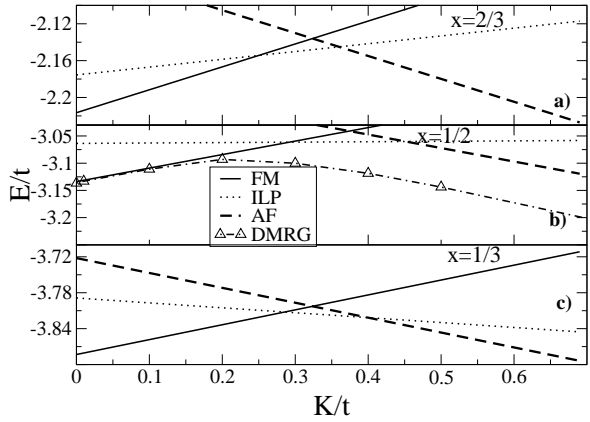}
\end{center}
\caption{Classical energies for different fillings in the FKLM. \label{clasicos}}
\end{figure}

The validity of the picture described above can be checked in the classical limit
for the localized spins. We calculated the contributions of DE and SE to the energy of the different 
possible phases for each concentration $x$. To this end, we assumed that the hopping at the AF bonds is $t_{AF}=t/\sqrt{2}$
(the conclusions are the same if we take $t_{AF}=0$). This value of  $t_{AF}$ gives a better agreement for the charge distribution  
in the ILP states and with the critical values of $K/t$ separating different phases.
For instance, for {\bf \ }$x=1/2$ we obtained $E_{FM}=-2t/\pi +K/4-J_H/8$ , $E_{\pi /2}=-.564761t-J_H/8$, 
and $E_{AF}=-\sqrt{2}t/\pi-K/4-J_H/8$. In this way, the sequence of stable spin phases as
a function of 
increasing $K/t$, FM$\rightarrow \pi /2\rightarrow $AF, coincides with our numerical results for quantum spins
(see Figs. \ref{phasediagram} and \ref{clasicos}b).
This simple calculation  clearly shows  how the FM state weakens by increasing $K/t$ and gives rise to more 
complicated structures when both interactions are comparable.
In this classical picture, a canted AF phase has a slightly lower energy than the pure AF phase. However the 
$\pi /2$ island phase remains  as the most stable state over a finite interval in $K/t$.
The same procedure leads to similar conclusions for $x=2/3$ and $1/3$ (see Figs. \ref{clasicos}a and c).

\subsection{\textit{Magnetic mechanism for pairing}}

As K is further increased ($K\gtrsim t$ and $V=0$), we find a peak at $q=\pi/4$ in the charge structure factor(Fig. \ref{distribcarga1s2}).
To understand the origin of this peak we will consider
the effective Hamiltonian which emerges by taking the limit of infinite $J_H$.
In this limit, the original Hamiltonian $H$ (Eq. (1)) can be projected into the subspace 
having finite energy, i.e., each carrier represented by an effective spin $S=1$  
and each hole represented by a spin $S=1/2$. 
Double occupation is projected out for infinite $J_H$. The projected Hamiltonian can be written as:

\begin{eqnarray}
\label{heff}
  H_{eff}&=&t \sum_{<i,j>} P_{i,j} (2 n_i n_j- n_i- n_j)({\bf S}_{i}{\bf S}_{j}+\frac
 {1}{2}) \nonumber \\ \nonumber &+&K \sum_{<i,j>}
[(1-n_i)(1-n_j) ({\bf S}_{i}{\bf S}_{j}-\frac
 {1}{4}) \\ \nonumber &+& \frac{1}{2} (2 n_i n_j- n_i- n_j) ({\bf S}_{i}{\bf S}_{j}-\frac
 {1}{2}) \\  &+& \frac{2}{9} n_i n_j ({\bf S}_{i}{\bf S}_{j}-1)] 
\end{eqnarray}
where $i,j$ are nearest neighbors and $n_i$ is the occupation number at site $i$. 
$n_i=0$ if there is a spin $S=1/2$ at site $i$ (hole) and $n_i=1$ if there is a spin $S=1$ (one conduction electron).
$P_{i,j}$ is a permutation operator between the spins at sites $i$ and $j$.
Note that this permutation is operating only between spins 1 and 1/2
due to the factor $(2 n_i n_j- n_i- n_j)$.
This is because the hopping process is only possible if the occupation numbers at sites $i$ and $j$ are different.
The last three terms of Eq. (\ref{heff}) are derived from the original AF interaction $K$ between localized spins.
This interaction remains the same between two spins $S=1/2$, but the situation is different between a spin $S=1$ and a spin $S=1/2$, or between two spins $S=1$.
In the last two cases the effective AF interaction is reduced due to the fact that the localized spin $S=1/2$ which is part of a spin $S=1$ cannot form a singlet state with its nearest neighbor spin. 
While the minimum AF energy for the singlet state between two spins $S=1/2$ (2 holes) is $-K$, the 
corresponding energy for the doublet between a spin $S=1$ and a spin $S=1/2$ (one particle and one hole) 
is $-\frac{3K}{4}$. The singlet between two spins $S=1$ (two particles) has an AF energy  $-\frac{2K}{3}$. 
The renormalization of the AF interaction gives rise to an attractive interaction for the charge degrees 
of freedom. The magnitude of this interaction is given by $V_{eff}=E_{2part}+E_{2holes}-2E_{1part}=-\frac{2K}{3}-K+\frac{3K}{2}=-\frac{K}{6}$.
It is well known that a nearest-neighbor attraction gives rise to three different phases depending on the 
ratio between the attractive interaction $V_{eff}$ and the effective hopping ($\sim t/2$). If this ratio is smaller than some 
critical value ($\sim \sqrt{2}$) the system is a Luttinger liquid, and the metallic correlations are the dominating ones \cite{tjz}.
Above that value the system remains as a Luttinger liquid but the superconducting correlations become dominant. 
It is precisely in this region where the formation of pairs gives rise to a peak at $q=\pi/4$ in the charge structure factor
for half filling. If the attractive interaction $V_{eff}$ is larger than the double of the effective hopping ($V_{eff}>t$),
the system segregates into two different phases. One phase is rich in spins $S=1/2$ (holes) and  the other one 
is rich in spins $S=1$ (electrons).  
A first indication of this behavior is presented in \cite{riera}, where binding and phase segregation is obtained in a similar 
model. However the binding energy is overestimated because the authors
have only included the AF interaction between spins $S=1/2$, i.e., they did not renormalized the AF effective interaction.

\section{Metallic and Insulating phases}

The diversity of phases found for the FKLM suggests different transport properties. 
To determine the metal or insulator character of the phases we have calculated the charge gap and the Drude weight ($V=0$) for $x=1/3, 1/2$, and $2/3$.

The charge gap for systems having $N$ particles is defined as:

\begin{equation}
\label{eqGap}
\Delta _{c}=E_{N+1}+E_{N-1}-2E_{N}.
\end{equation}
\( E_{N} \) is the ground state energy of the considered system.
\( E_{N+1} \) and \( E_{N-1} \) are the ground states for $N+1$ and $N-1$ particles respectively. 
To get the charge gap in the thermodynamic limit we did a $1/L$ extrapolation of the gap $\Delta _{c}(L)$ for different sizes.

The Drude weight $D(L)$ is defined as:
\begin{equation}
\label{EqDrude}
2\pi D(L)=\frac{1}{L}\frac{\partial ^{2}E}{\partial \phi ^{2}}
\end{equation}
where \( \phi  \) can be interpreted as the magnetic flux inside a closed chain. 
$D(L)$ was calculated by exact diagonalization in closed chains having 8 and 12 sites.
The ground state used to compute $D(L)$ is the one which
minimizes the energy as a function of the flux $\phi$.
To reduce the Hilbert space, we did this calculation with the Hamiltonian (\ref{hamReducido}).
The results were tested  by doing the same calculation with the full Hamiltonian (\ref{Hamiltoniano}) 
in a half filled chain with 8-sites. We did not find any substantial variation.   

\begin{figure}
\begin{center}
\includegraphics[height=5.cm]{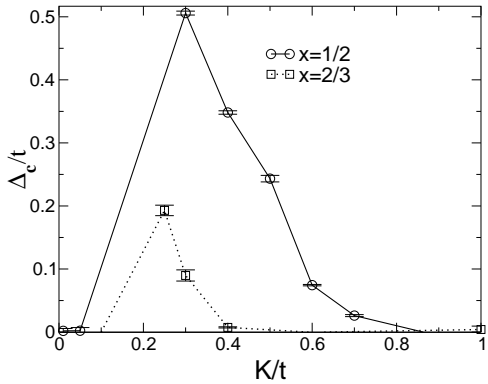}
\end{center}
\caption{Extrapolated charge gap in the thermodynamic limit. The lines are only a guide to the eye.\label{gapcarga}}
\end{figure}
\begin{figure}
\begin{center}
\includegraphics[height=5.cm]{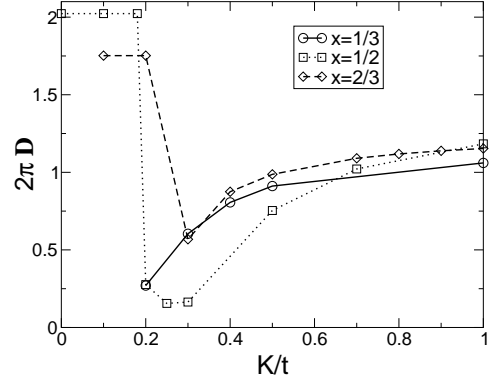}
\end{center}
\caption{Drude weight for different fillings, $J/t=0.4$, $V=0$  and $L=12$ sites as a function of the super-exchange.\label{Drudeweight}}
\end{figure}

In the FM phase, the charge gap is zero as it is shown in Fig. \ref{gapcarga}. Consistently with the cancellation of the  gap, there 
is a finite Drude weight which coincides with the one obtained for a spinless model 
(see Fig. \ref{Drudeweight}). When $K/t$ is large enough to get into the ILP,  $D(L)$ decreases abruptly and approaches zero. 
This is indicative of a metal-insulator transition. These results are not conclusive because
we cannot solve large enough chains to make an accurate extrapolation to the thermodynamic 
limit. Another indication of the insulating character of the ILP state for $x=1/2$ and $2/3$ is the 
finite value of the charge gap. The gap for $x=1/2$ is greater than the one for $x=2/3$, as it is
expected from a simple calculation with classical spins. For $x=1/3$, the gap calculated in the ILP is small 
($\Delta_{c} < 0.05t$), but the precision does not allow us to obtain a reliable extrapolation to the thermodynamic limit. 
$D$ increases for larger values of K, but it is unclear whether the AF state is insulating or metallic. 
The charge gap decreases and cancels when the AF phase is established (Fig. \ref{gapcarga}). 
The closure of the charge gap and the finite value of D suggest a metallic character for this phase.
\\

\begin{figure}
\begin{center}
\includegraphics[height=5.cm]{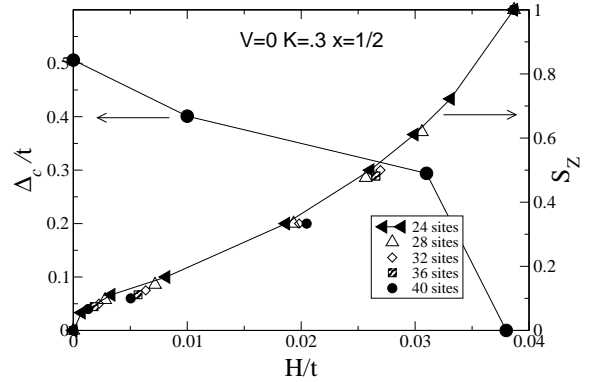}
\end{center}
\caption{Mean on-site magnetization $S_z$ and extrapolated charge gap ($\Delta_{c}$) as a function of the magnetic field (H) in the ILP phase for x=1/2.\label{transicionconB}}
\end{figure}

To investigate the possibility of an insulator-metal transition with magnetic field applied to the ILP phase, 
we have calculated the lowest energy state for different magnetizations ($L\leq40$). 
For $x=1/2$ we found that the total spin increases smoothly with field and the charge gap closes abruptly when the 
system reaches its maximum magnetization (see Fig. \ref{transicionconB}). A similar behavior is seen in the ILP of $x=2/3$.
As it is expected, the field needed to induce the transition from the ILP to the FM phase decreases when $K/t$
gets close to the critical value. This is shown in Fig. \ref{clasicos}, where we compare the energies of the different phases.

\section{Conclusions}

In summary, we have presented numerical evidence for the existence of a new type of charge and 
spin ordering in the FKLM. This mechanism, induced by the competition between double and super-exchange, is 
based on a striking interplay between charge and spin degrees of freedom.
As we have shown in a previous work \cite{Nos1}, in the dilute limit each carrier polarizes its surroundings 
inducing a FM distortion. The quasiparticles constituted by the electron plus the magnetic 
distortion are called FM polarons. The size of these polarons is governed by the ratio \( K/t \). When
the mean separation between carriers equals the size of the FM distortion associated to each polaron (FM islands), 
the SE interaction induces an AF ordering  between polarons.
This is of course a schematic picture since there are quantum magnetic fluctuations within and between the 
islands. 

While previous works reported the need of a Coulomb interaction \( V \) \cite{malvezzi} or electron-phonon 
coupling \( \lambda  \) \cite{yunokihotta} to stabilize a charge-ordered phase, we show that it arises naturally 
just by considering the competition between double and super-exchange.

A great variety of experiments have found simultaneous charge and spin ordering in manganites. 
The extraordinary colossal magnetoresistance effect for La\( _{0.5}\)Ca\( _{0.5} \)MnO\( _{3} \)
involves the abrupt destabilization of a CO-AF state by application of a magnetic field \cite{Tomioka}.
Insulating charge-ordered and metallic FM regions coexist 
in (La\( _{0.5} \)Nd\( _{0.5} \))\( _{2/3} \)Ca\( _{1/3} \)MnO\( _{3} \)
\cite{Ibarra} and Pr\( _{0.7} \)Ca\( _{0.3} \)MnO\( _{3} \) \cite{Cox,Kiryukhin}.
Both phenomena indicate that the CO phase is very close in energy to the FM state.
We have shown here that a magnetic field can produce an abrupt metal-insulator transition by polarizing the local spins and favouring the double exchange mechanism.
For these reasons, we think that the phenomenon presented here is relevant for the underlying physical mechanism stabilizing spin and charge structures in manganites. 


Our model calculations fail to include several effects that may play an important role in real systems like Jahn-Teller distortions and orbital degeneracy. 
However, the spirit of the present paper is to show that a minimal model, including the most relevant electron-electron interactions, is sufficient to explain the simultaneous appearing of charge and spin ordering.
This is a possibility which had been excluded by the spin canted phases proposed in previous works \cite{de gennes}.

\section*{acknowledgments}

We thank S. Bacci for helping us with the numerical calculations, and 
J. Gubernatis for a careful reading of the manuscript. K. H. and
D. J. G. are supported by CONICET. B. A. is partially suported by CONICET.
C.D.B. is supported by Los Alamos National Laboratory. Work at Los Alamos 
is sponsored by the US DOE under contract W-7405-ENG-36. Support from the cooperation program between 
France and Argentina ECOS-SETCIP A97E05 is greatly acknowledged.

\pagebreak

\newpage

\end{multicols} 

\end{document}